\documentclass{appolb}
\usepackage{epsfig}
\newcommand\beq{\begin{eqnarray}}
\newcommand\eeq{\end{eqnarray}}
\newcount\eLiNe\eLiNe=\inputlineno\advance\eLiNe by -1
\begin{document}
\title{CP VIOLATION IN ANTINEUTRINO-ELECTRON ELASTIC SCATTERING}
\author{W. SOBK\'OW  \and S. CIECHANOWICZ 
\address{Institute of Theoretical Physics, University of Wroc\l{}aw,
Pl. M. Born 9, PL-50-204~Wroc{\l}aw, Poland\\
e-mail: {\tt ciechano@rose.ift.uni.wroc.pl}  \\ e-mail: {\tt
sobkow@rose.ift.uni.wroc.pl}}
 \and  M. MISIASZEK
 \address{M. Smoluchowski
Institute of Physics, Jagiellonian University, ul. Reymonta 4,\\
PL-30-059 Krak\'ow, Poland \\
 e-mail: {\tt misiaszek@zefir.if.uj.edu.pl}}}
\maketitle
\begin{abstract}
In this paper we show that  the elastic scattering of transversely polarized electron
antineutrino beam off unpolarized electrons can be used to detect the CP-violating effects by measuring the  azimuthal asymmetry of recoil electrons caused by the interference terms 
between the standard vector $c_{V}^{L}$, axial $c_{A}^{L}$ couplings of left-chirality 
antineutrinos and exotic scalar $c_{S}^{R}$  coupling of 
right-chirality ones  in the differential cross section. It would be a positive evidence for the existence of the exotic antineutrino states. 
Moreover, we also show that the differential cross section for the $\overline{\nu} e^- $ scattering can 
be obtained from the one for the $\nu  e$ scattering, if one makes the substitution $c_{T}^{R} \rightarrow -c_{T}^{R}$, $c_{A}^{L} \rightarrow -c_{A}^{L}$, ${\bf q}\rightarrow -{\bf q}$, $\mbox{\boldmath $\hat{\eta}_{\nu}$}\rightarrow -\mbox{\boldmath $\hat{\eta}_{\nu}$}$.    
Electron antineutrinos are assumed to be massive  and to be
polarized Dirac fermions coming from the polarized muon decay at rest. The
results are presented in a limit of infinitesimally small antineutrino mass.  
\end{abstract}


\section{Introduction}
Neutrino-electron elastic scattering is a suitable process to test the CP violation,  Lorentz structure, chiral structure and most of all possibility of participation of the right-chirality neutrinos in the purely leptonic charged and neutral weak interactions. In addition, $\nu e$ scattering can also be used to proble if existing neutrinos are Dirac or Majorana fermions. \\
As is well known, the CP violation is observed only in the decays of neutral kaons and B-mesons \cite{CP} and is described by a single phase of the CKM matrix \cite{Kobayashi}. However, the baryon asymmetry of the universe can not be explained by the standard CKM phase only, and new sources of breaking CP symmetry is needed \cite{Riotto}. It is worth to notice that there is no proof of the CP violation in the leptonic processes, i. e. in (anti)neutrino-electron scatteing or muon decay. \\
F. Wilczek et al. \cite{Wilczek} considered the $\nu e$ scattering and searched for the effects of  anomalous (nonstandard) Lorentz structure in the weak interactions. They admitted the most general local (derivative free) Lagrangian including the five types of Lorentz covariants; scalar, pseudoscalar, tensor, vector and axial-vector.  In their case, the incoming neutrinos was always left-chirality and longitudinally polarized. In consequence, no interference terms between the standard and nonstandard couplings was present in the differential cross section. They also showed that the differential cross section for the $\overline{\nu} e^- $ scattering can 
be obtained from the one for the $\nu  e$ scattering, if one simply substitutes $c_{T} \rightarrow -c_{T}$, $c_{A} \rightarrow -c_{A}$.\\ Cheng and Tung \cite{Cheng} proposed   the measurement of polarization of the outgoing lepton in the framework of general local current-current interaction assuming only left-chirality incoming neutrinos. They pointed out that such measurement would allow to  distinguish the V-A interactions from the other S, T, P admixtures. Since the direct tests would be  extremely difficult at very high energies, they indicated the angular and spin correlation experiments for testing the Lorentz structure. This proposal involved the measurement of the outgoing lepton angular distribution in either the differential cross section or in certain asymmetry functions.\\
T.C. Yang \cite{Yang} probed the $\nu  e$ processes in which the right-chirality neutrinos,  produced in the exotic S, T, P weak interactions,  can take part and calculated the angular distribution of the scattered electrons for an unpolarized and a polarized electron target. He showed that if the right-chirality neutrinos are present in the incoming neutrino beam and $\nu_R  e \rightarrow \nu_L  e$ scattering  occurs, their effect should be seen in near backward directions of the electron in the c. m. system. 
\par In this paper, we study  the elastic scattering of transversely polarized  electron 
antineutrino beam on the unpolarized electron target and predict the new effects beyond the standard model of electroweak interactions \cite{SM}.  The main goal is to show that  the CP violation in the scattering of a mixture of the standard left-chirality electron antineutrinos and exotic right-chirality  ones on the unpolarized electron target can be observed due to  the azimuthal asymmetry of recoil electrons. In addition, we show how to get the differential cross section for the $\overline{\nu} e^- $ scattering from the corresponding one for the case with incoming neutrinos. 
\par
 We use the system of natural units with $\hbar=c=1$, Dirac-Pauli
representation of the $\gamma$-matrices and the $(+, -, -, -)$
metric \cite{Mandl}.

\section{Basic assumptions}
We consider the $\overline{\nu} e^- $ scattering, when the incoming electron antineutrino beam is a mixture the left-chirality electron antineutrinos produced in the standard vector charged weak interaction and the right-chirality ones produced in the exotic scalar charged weak interaction. This beam comes from the polarized muon decay  at rest and has a nonzero value of the  transverse antineutrino spin polarization $ \mbox{\boldmath
$\eta_{\overline{\nu}}^\perp$}$. A direction of  this transversal polarization is assigned with respect to the production plane spanned by the direction of initial muon polarization $\mbox{\boldmath $\hat{\eta}_{\mu}$}$ and of the outgoing antineutrino momentum ${\bf \hat{q}}$. The reaction plane is spanned by the direction of  antineutrino momentum ${\bf \hat{q}} $ and of the outgoing electron momentum $ \hat{\bf p}_{e}$, see Fig.1. As is known,  the polarization vector $\mbox{\boldmath $\hat{\eta}_{\mu}$}$  can
be expressed, with respect to the $\hat{\bf q}$, as a sum of the longitudinal
component of the muon polarization
$(\mbox{\boldmath$\hat{\eta}_{\mu}$}\cdot\hat{\bf q}){\bf \hat{q}} $
 and transverse component of the muon polarization 
$\mbox{\boldmath $\eta_{\mu} ^{\perp}$} $, that is defined as 
$ \mbox{\boldmath $\eta_{\mu}^{\perp}$} =
\mbox{\boldmath$\hat{\eta}_{\mu}$}-
(\mbox{\boldmath$\hat{\eta}_{\mu}$}\cdot\hat{\bf q}) {\bf \hat{q}}$.
\\
\begin{figure}
\label{pmu}
\begin{center}
\includegraphics*[scale=.5]{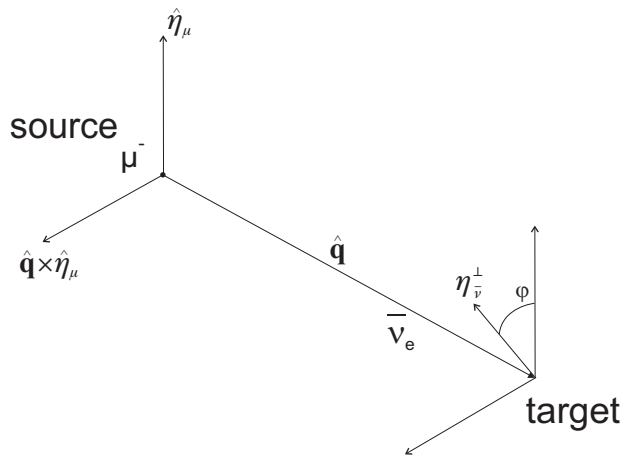}
\end{center}
\caption{Figure shows the production plane of the
transversely polarized electron antineutrino beam in $\mu^{-} \rightarrow e^-
+ \overline{\nu}_{e} + \nu_{\mu}$ and  the reaction plane in $\overline{\nu} e^- $ scattering.}
\end{figure}
We assume that the left-chirality antineutrinos are detected in the standard $V-A$ charged interaction with the unpolarized electrons, while the right-chirality ones are detected in the exotic scalar  one. In the limit o vanishing antineutrino mass, the left-chirality antineutrino has a positive helicity, while the right-chirality one has a negative helicity.
We want to show how the differential cross section for the $\overline{\nu} e^- $ scattering depends on the CP-violating relative phase between the standard vector and exotic scalar couplings.
We also assume that a detector is able to measure both the recoil electron scattering angle and the azimuthal angle of outgoing electron momentum with a high angular resolution.
Because we
allow for the nonconservation of the combined symmetry CP, the
transition amplitude includes the complex coupling constants denoted as $c_{V}^{L}, c_{A}^{L}$ and $c_{S}^{R}$ respectively to the initial electron antineutrino of left- and right-chirality: 
\beq \label{amp} M_{\overline{\nu} e}
&=&
\frac{G_{F}}{\sqrt{2}}\{(\overline{u}_{e'}\gamma^{\alpha}(c_{V}^{L}
- c_{A}^{L}\gamma_{5})u_{e}) (\overline{v}_{\nu_{e}}
\gamma_{\alpha}(1 - \gamma_{5})v_{\nu_{e'}})\nonumber \\ &  &
\mbox{} +
\frac{1}{2}c_{S}^{R}(\overline{u}_{e'}u_{e})(\overline{v}_{\nu_{e}}
(1 - \gamma_{5})v_{\nu_{e'}})\},
 \eeq
 where $ u_{e}$ and  $\overline{u}_{e'}$
$(\overline{v}_{\nu_{e}}\;$ and $\; v_{\nu_{e'}})$ are the
Dirac bispinors of the initial and final electron (electron antineutrino)
respectively. $G_{F}= 1.16639(1)\times 10^{-5}\,\mbox{GeV}^{-2}$
 is the Fermi constant. $c_{V}^{L}= -0.040+1, c_{A}^{L}=-0.507+1$ \cite{Data}.\\
An admittance of tensor and psudoscalar couplings of the right-chirality antineutrinos does not change qualitatively the
conclusions from the studies.  In addition, if the incoming antineutrino beam consists only of the  left-chirality antineutrinos produced in the standard and exotic weak interactions, there is no interference
between the  $c_{V, A}^{L}$ and $c_{S}^{L} $ couplings in the
differential cross section, when $m_{\nu}\rightarrow 0$. We do not
consider this scenario.  

\section{Azimuthal asymmetry of recoil electrons}

 The laboratory differential cross
section for the $\overline{\nu} e^- $ scattering, in the limit of
vanishing antineutrino mass, has  the form:

 \beq \label{przekranue} \frac{d^{2}
\sigma}{d y_{e} d \phi_{e}} &=& \bigg( \frac{d^{2} \sigma}{d y_{e} d
\phi_{e}}\bigg)_{(V, A)} + \bigg( \frac{d^{2} \sigma}{d y_{e} d \phi_{e}}\bigg)_{(S)} 
 + \bigg( \frac{d^{2} \sigma}{d y_{e} d \phi_{e}} \bigg)_{(V S)}, 
%+ \bigg( \frac{d^{2}\sigma}{d y_{e} d \phi_{e}} \bigg)_{(A T)}, \nonumber  
\\ \bigg( \frac{d^{2} \sigma}{d
y_{e} d \phi_{e}} \bigg)_{(V, A)} &=& \frac{E_{\nu}m_{e}}{4\pi^2} \frac{G_{F}^{2}}{2} \bigg\{ (1+\mbox{\boldmath
$\hat{\eta}_{\overline{\nu}}$}\cdot\hat{\bf q}
) \bigg[(c_{V}^{L} - c_{A}^{L})^{2} \\
&& \mbox{} + (c_{V}^{L}+ c_{A}^{L})^{2}(1-y_{e})^{2}
 - \frac{m_{e}y_{e}}{E_{\nu}}\left((c_{V}^{L})^{2} - (c_{A}^{L})^{2}\right) \bigg] \bigg\},
\nonumber \\ \bigg(\frac{d^{2} \sigma}{d y_{e} d \phi_{e}}\bigg)_{(S)} &=& \mbox{}
\frac{E_{\nu}m_{e}}{4\pi^2} \frac{G_{F}^{2}}{2} (1-\mbox{\boldmath $\hat{\eta}_{\overline{\nu}}$}\cdot\hat{\bf
q})\bigg\{\frac{1}{8}y_{e}\left(y_{e}+2\frac{m_{e}}{E_{\nu}}\right)
  |c_{S}^{R}|^{2} \bigg\},\\   
%\\&& \mbox{} + \left((2-y_{e})^2 -\frac{m_{e}}{E_{\nu}}y_{e}\right)|{c_{T}^{R}}|^{2}
%- \frac{1}{2}y_{e}(y_{e}-2)Re(c_{S}^{R}c_{T}^{*R}) \bigg\}, \nonumber\\
  \bigg(\frac{d^{2}\sigma}{d y_{e} d \phi_{e}}\bigg)_{(V S)} &=& \mbox{}
\frac{E_{\nu}m_{e}}{4\pi^2} \frac{G_{F}^{2}}{2}  \bigg\{ -\sqrt{y_{e}(y_{e}+2\frac{m_{e}}{E_{\nu}})}\bigg[(\mbox{\boldmath
$\eta_{\overline{\nu}}^{ \perp}$}\cdot {\bf \hat{p}_{e}}) Re(c_{V}^{L}c_{S}^{R*})
\nonumber \\ && \mbox{} + \mbox{\boldmath
$\eta_{\overline{\nu}}^{ \perp}$}\cdot({\bf \hat{p}_{e} \times
\hat{q}})Im(c_{V}^{L}c_{S}^{R*}) \bigg]
\bigg\},
%\\ \bigg( \frac{d^{2} \sigma}{d y_{e} d \phi_{e}}\bigg)_{(A T)} &=& B \bigg\{ 
%-\sqrt{y_{e}(y_{e}+2\frac{m_{e}}{E_{\nu}})}\bigg[\mbox{\boldmath
%$\eta_{\overline{\nu}}^{ \perp}$}\cdot({\bf \hat{p}_{e} \times
%\hat{q}})Im(c_{A}^{L}c_{T}^{R*}) \\ && \mbox{} + (\mbox{\boldmath
%$\eta_{\overline{\nu}}^{ \perp}$}\cdot {\bf \hat{p}_{e}}) Re(c_{A}^{L}c_{T}^{R*})\bigg]
%\bigg\},\nonumber 
\eeq
\beq  y_{e} & \equiv &
\frac{T_{e}}{E_{\nu}}=\frac{m_{e}}{E_{\nu}}\frac{2cos^{2}\theta_{e}}
{(1+\frac{m_{e}}{E_{\nu}})^{2}-cos^{2}\theta_{e}} \eeq is the ratio of the
kinetic energy of the recoil electron $T_{e}$  to the incoming (anti)neutrino
energy $E_{\nu}$; 
$\theta_{e}$ is the angle between the direction of the outgoing electron
momentum  $ \hat{\bf p}_{e}$  and the direction  of the incoming (anti)neutrino
momentum $\hat{\bf q}$ (recoil electron scattering angle); $m_{e}$ is the
electron mass; 
%$ \mbox{\boldmath $\hat{\eta}_{\overline{\nu}}$}\cdot\hat{\bf q}$
%is the longitudinal polarization of the incoming antineutrino; 
$\phi_{e}$ is the
angle between the production plane and the reaction plane spanned by the $
\hat{\bf p}_{e}$ and $ \hat{\bf q}$.
\\
We see that the term with interference  between the
standard $c_{V}^{L}$ and exotic $c_{S}^{R}$ couplings does not depend on the antineutrino mass, so does not vanish in the limit of vanishing antineutrino mass. It is proportional to 
the transverse components of the initial antineutrino spin  polarization, both $CP$-even
and $CP$-odd. This interference can be rewritten as follows: \beq \label{intervs} \bigg(\frac{d^{2} \sigma}{d y_{e} d
\phi_{e}}\bigg)_{(VS)}  &=& -\frac{E_{\nu}m_{e}}{4\pi^2} \frac{G_{F}^{2}}{2}
|\mbox{\boldmath $\eta_{\overline{\nu}}^{ \perp}$}|
\sqrt{\frac{m_{e}}{E_{\nu}}y_{e}[2-(2+\frac{m_{e}}{E_{\nu}})y_{e}]}\\
&& \mbox{} \cdot \bigg[ |c_{V}^{L}||c_{S}^{R}| cos(\phi-\alpha_{SV}-\phi_e)\bigg]\nonumber . 
%& + & |c_{A}^{L}||c_{T}^{R}|cos(\phi-\alpha_{TA}-\phi_e)\}, \nonumber 
\eeq where
$\alpha_{SV} \equiv\alpha_{S}^{R} - \alpha_{V}^{L}$, 
%$\alpha_{TA}\equiv\alpha_{T}^{R} - \alpha_{A}^{L} $
- the relative phase between the $c_{S}^{R}$, $c_{V}^{L}$ couplings. 
% and $c_{T}^{R}$, $c_{A}^{L}$  respectively. 
The interference contribution is linear in the $c_{S}^{R}$ coupling and contains the relative phase $\alpha_{SV}$ which could generate the CP violation, when $\alpha_{SV}\not = 0$ 
or $\pi$. 
The appearance of above interference in the cross section should manifest the observation of azimuthal asymmetry of the scattered electrons. This asymmetry does vanish even for $\alpha_{SV}= 0$, however there is  different azimuthal dependence in th case of CP conservation and CP nonconservation. 
It is necessary to point out that with the proper choice of angle $\phi $, a measurement of the maximal asymmetry of the cross section could detect the CP-violating phase.  
\begin{figure}
\label{pmu}
\begin{center}
\includegraphics*[scale=.9]{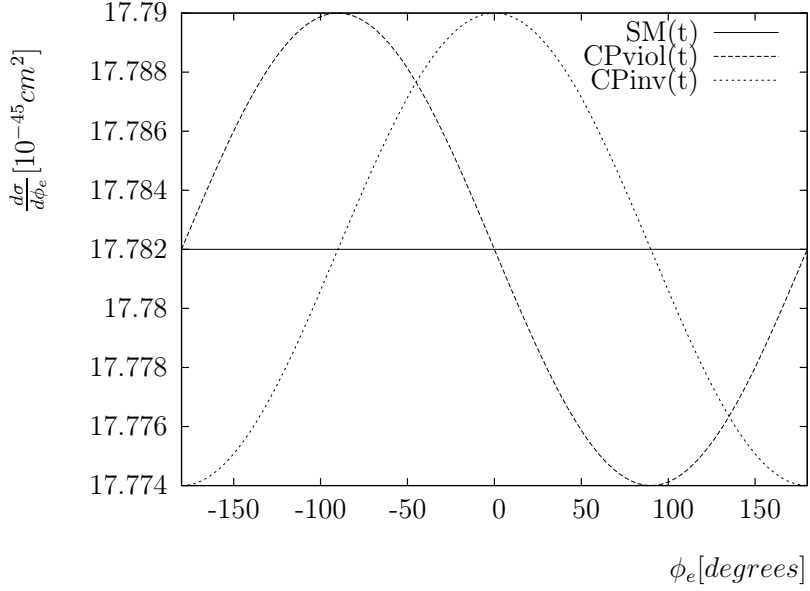}
\end{center}
\caption{Plot of the $\frac{d
\sigma}{d \phi_{e}}$ as a
function of the $\phi_{e}$ for 
%\mbox{\boldmath$\hat{\eta}_{\mu}$}\cdot\hat{\bf q}=0,
$\mbox{\boldmath$\hat{\eta}_{\overline{\nu}}$}\cdot\hat{\bf q}=0.996,
|\mbox{\boldmath $ \eta_{\overline{\nu} }^{\perp}$}|=0.088,  y_{e}=1/2, |c^{R}_{S}|=0.088, 
|c^{L}_{V}|=0.96, |c^{L}_{A}|=0.493$. The solid line is for the SM case, the long-dashed line corresponds to 
the CP violation for $\alpha_{VS}=\pi/2$, while the short-dashed line  represents  the CP symmetric case  for the $\alpha_{VS}=0$.}
\end{figure}
The Fig.2 shows the possble effect of the CP violation connected with the interference term $c_{V}^{L}c_{S}^{R*}$ proportional to the $|\mbox{\boldmath $\eta_{\overline{\nu}}^{ \perp}$}|$. 
\\ 
To give a numerical example of  expected event number, we assume that in our analysis an   antineutrino source is located in the center of the ring detector and is polarized perpendiculary to the ring. Moreover, we assume that $T_{e}^{th}=100 keV $ - a detector threshold; $N_{e}= 2.097 \cdot 10^{34}$ - number of electrons in fiducial volume of the detector; $\epsilon=1$ - an efficiency of the detector for antineutrino energy above threshold; $N_\mu= 10^{20}$ - number of muons decaying per year. This number gives the antineutrino flux, i.e. the number of antineutrinos passing  through $S_{D}=2 \pi R \cdot D=305490 cm^{2}$ (where $R=L=2205 cm$ is the inner radius of the ring that is equal to the distance from the antineutrino source, $D=22.05 cm$ is the thickness of the ring detector) in the direction perpendicular to the $\mbox{\boldmath $\hat{\eta}_{\mu}$}$ according to the SM: $\Phi_{\overline{\nu}}^{\perp}= 1.497\cdot 10^{18} cm^{-2}s^{-1}$.  For the SM, the event number does not depend on the $\phi_{e}$ and one expects $\frac{dN_{e}}{d\phi_{e}} \simeq 1.52 \cdot 10^8$ events (recoil electrons) per year. To calculate the event number we used the antineutrino spectral function, see Appendix B. 
 If the  exotic $c_{S}^{R}$ coupling is present in $\overline{\nu} e^- $ scattering, the azimuthal asymmetry of the event number should occur.   
 \\It is worth to notice that a knowledge of the differential cross section for $\nu e^- $ scattering allows to get the correct formula for $\overline{\nu} e^- $ scattering, making the simple substitutions.
 The calculated formula for the laboratory differential cross
section in the case of  $\nu e^- $ scattering, in the limit of
vanishing neutrino mass, is as follows:
 \beq \label{przekrnue} \frac{d^{2}
\sigma}{d y_{e} d \phi_{e}} &=&  \bigg( \frac{d^{2} \sigma}{d y_{e} d
\phi_{e}}\bigg)_{(V, A)} + \bigg( \frac{d^{2} \sigma}{d y_{e} d \phi_{e}}\bigg)_{(S)} 
 + \bigg( \frac{d^{2} \sigma}{d y_{e} d \phi_{e}} \bigg)_{(V S)}, 
%+ \bigg( \frac{d^{2}\sigma}{d y_{e} d \phi_{e}} \bigg)_{(A T)}, \nonumber  
\\ \bigg( \frac{d^{2} \sigma}{d
y_{e} d \phi_{e}} \bigg)_{(V, A)} &=& \frac{E_{\nu}m_{e}}{4\pi^2} \frac{G_{F}^{2}}{2} 
 \bigg\{ (1-\mbox{\boldmath $\hat{\eta}_{\nu}$}\cdot\hat{\bf q}
) \bigg[(c_{V}^{L} + c_{A}^{L})^{2} \\ && \mbox{} + (c_{V}^{L}- c_{A}^{L})^{2}(1-y_{e})^{2}
 - \frac{m_{e}y_{e}}{E_{\nu}}\left((c_{V}^{L})^{2} - (c_{A}^{L})^{2}\right) \bigg] \bigg\},
\nonumber \\ 
\bigg(\frac{d^{2} \sigma}{d y_{e} d \phi_{e}}\bigg)_{(S)} &=& \mbox{}
\frac{E_{\nu}m_{e}}{4\pi^2} \frac{G_{F}^{2}}{2} (1+\mbox{\boldmath $\hat{\eta}_{\nu}$}\cdot\hat{\bf
q})\bigg\{\frac{1}{8}y_{e}\left(y_{e}+2\frac{m_{e}}{E_{\nu}}\right)
  |c_{S}^{R}|^{2}  \bigg\},  \\
%&& \mbox{} + \left((2-y_{e})^2 -\frac{m_{e}}{E_{\nu}}y_{e}\right)|{c_{T}^{R}}|^{2}
%+ \frac{1}{2}y_{e}(y_{e}-2)Re(c_{S}^{R}c_{T}^{*R}) \bigg\}, \nonumber\\
  \bigg(\frac{d^{2}\sigma}{d y_{e} d \phi_{e}}\bigg)_{(V S)} &=& \mbox{}
\frac{E_{\nu}m_{e}}{4\pi^2} \frac{G_{F}^{2}}{2} \bigg\{ \sqrt{y_{e}(y_{e}+2\frac{m_{e}}{E_{\nu}})}\bigg[-\mbox{\boldmath
$\eta_{\nu}^{ \perp}$}\cdot({\bf \hat{p}_{e} \times
\hat{q}})Im(c_{V}^{L}c_{S}^{R*}) \nonumber \\ && \mbox{} + (\mbox{\boldmath
$\eta_{\nu}^{ \perp}$}\cdot {\bf \hat{p}_{e}}) Re(c_{V}^{L}c_{S}^{R*})\bigg]
\bigg\}.
%\bigg( \frac{d^{2} \sigma}{d y_{e} d \phi_{e}}\bigg)_{(A T)} &=& B \bigg\{ 
%\sqrt{y_{e}(y_{e}+2\frac{m_{e}}{E_{\nu}})}\bigg[-\mbox{\boldmath
%$\eta_{\nu}^{ \perp}$}\cdot({\bf \hat{p}_{e} \times
%\hat{q}})Im(c_{A}^{L}c_{T}^{R*}) \\ && \mbox{} + (\mbox{\boldmath
%$\eta_{\nu}^{ \perp}$}\cdot {\bf \hat{p}_{e}}) Re(c_{A}^{L}c_{T}^{R*})\bigg]
%\bigg\}.\nonumber 
\eeq
We see that the Eq. (\ref{przekranue}) can be obtained from the Eq. ( \ref{przekrnue}) by  substituting $c_{T}^{R} \rightarrow -c_{T}^{R}$, $c_{A}^{L} \rightarrow -c_{A}^{L}$, ${\bf q}\rightarrow -{\bf q}$, $\mbox{\boldmath $\hat{\eta}_{\nu}$}\rightarrow -\mbox{\boldmath $\hat{\eta}_{\overline{\nu}}$}$, so $\mbox{\boldmath
$\eta_{\nu}^{ \perp}$} \rightarrow - \mbox{\boldmath $\eta_{\overline{\nu}}^{ \perp}$}$. In addition, $\mbox{\boldmath $\hat{\eta}_{\nu}$}\cdot\hat{\bf q}$ changes the sign respectively to the definition of the density operator for the polarized neutrino and antineutrino, see Appendix A. 

\section{Conclusion}

We have shown that the scattering of the electron anineutrino beam, produced in the decays of polarized muons at rest, on the unpolarized electron target  can be used to measure the CP violation  in leptonic weak interactions. An appropriate observable to unambiguous test would be the observation of azimuthal asymmetry of recoil electrons generated by nonzero  interference terms between the standard $c_{V}^{L}$ and exotic $c_{S}^{R}$ couplings, proportional to the magnitute of $\mbox{\boldmath $\eta_{\overline{\nu}}^{ \perp}$}$. \\
According to the standard model, the angular distribution of scattered electrons should be symmetric. The detection of the azimuthal asymmetry would indicate the possible existence of the right-chirality antineutrino states (it means that in this case   right-chirality antineutrinos have negative helicity when for $m_{\nu}\rightarrow 0$). \\ 
It is worth to point out that if the incoming anineutrino beam consists  only of the left-chirality and longitudinally polarized anineutrinos, there is no  interference $Re(c_{V}^{L} c_{S}^{L*})$ or $Im(c_{V}^{L} c_{S}^{L*})$ connected with CP violation, and the electron angular distribution is symmetric. This is in agreement with the Wilczek results. \\
We also have noticed a general regularity that a knowledge of the differential cross section for $\nu e^- $ scattering allows to write the corresponding formula for $\overline{\nu} e^- $ scattering,if one substitutes $c_{T}^{R} \rightarrow -c_{T}^{R}$, $c_{A}^{L} \rightarrow -c_{A}^{L}$, ${\bf q}\rightarrow -{\bf q}$, $\mbox{\boldmath $\hat{\eta}_{\nu}$}\rightarrow -\mbox{\boldmath $\hat{\eta}_{\overline{\nu}}$}$, so $\mbox{\boldmath
$\eta_{\nu}^{ \perp}$} \rightarrow - \mbox{\boldmath $\eta_{\overline{\nu}}^{ \perp}$}$, and one uses the correct definitions of the density operators for polarized antineutrino (neutrino). \\
The high-resolution neutrino-electron experiments require very large detectors and intense polarized neutrino sources which are very well understood (shape and normalization) to accumulate enough statistics because the cross section for $\nu e$ scattering is tiny. In addition, such experiments should run long (one year) and the detectors must distinguish the electrons from various potential background sources. New detectors should also  measure both the polar angle and the azimuthal angle of the outgoing electrons with high resolution. 
\appendix 

\section{Spin polarization 4-vector of  massive (anti)neutrino and density
operator of the polarized (anti)neutrino}

The formula for the  spin polarization 4-vector of  massive antineutrino
$ S^{\prime}{\overline{\nu}}$
%in its rest frame and for the initial neutrino
moving  with the momentum ${\bf q}$ is as follows:
%\begin{widetext}
\beq
S^{\prime}{\overline{\nu}} & = & (S^{\prime 0}_{\overline{\nu}}, {\bf S^{\prime}}_{\overline{\nu}}),\\
S^{\prime 0}_{\overline{\nu}} & = & \frac{{|\bf q|}}{m_{\overline{\nu}}}(\mbox{\boldmath
$\hat{\eta}_{\overline{\nu}}$}\cdot{\bf \hat{q}}),
\\
{\bf S^{\prime}}_{\overline{\nu}} & = &
-\left(\frac{E_{\overline{\nu}}}{m_{\overline{\nu}}}(\mbox{\boldmath
$\hat{\eta}_{\overline{\nu}}$}\cdot{\bf \hat{q}}){\bf \hat{q}} +
\mbox{\boldmath $\hat{\eta}_{\overline{\nu}}$} - (\mbox{\boldmath
$\hat{\eta}_{\overline{\nu}}$}\cdot{\bf \hat{q}}){\bf \hat{q}}\right),
\eeq
where $\mbox{\boldmath $\hat{\eta}_{\overline{\nu}}$}$ - the unit 3-vector of
the antineutrino polarization in its rest frame. The formula for the density
operator of the polarized antineutrino in the limit of vanishing  antineutrino
mass $m_{\overline{\nu}} $  is given by:
\beq
\lim_{m_{\overline{\nu}}\rightarrow 0}\Lambda_{\overline{\nu}}^{(s)} &=&
\lim_{m_{\overline{\nu}}\rightarrow 0}
\frac{1}{2}\bigg\{\left[(q^{\mu}\gamma_{\mu}) - m_{\nu}\right]\left[1 +
\gamma_{5}(S^{\prime \mu}_{\overline{\nu}}\gamma_{\mu})\right]\bigg\}  \\
%& = & \mbox{} \frac{1}{2}\bigg\{(q^{\mu}\gamma_{\mu})
%\left[1 - \gamma_{5}(\mbox{\boldmath
%$\hat{\eta}_{\overline{\nu}}$}\cdot{\bf \hat{q}}) + \gamma_{5} (\mbox{\boldmath
%$\hat{\eta}_{\overline{\nu}}$} - (\mbox{\boldmath $\hat{\eta}_{\overline{\nu}}$}\cdot{\bf
%\hat{q}}){\bf \hat{q}})\cdot
%\mbox{\boldmath $\gamma$}\right]\bigg\} \\
 & = & \mbox{} \frac{1}{2}\bigg\{(q^{\mu}\gamma_{\mu})
 \left[1 - \gamma_{5}(\mbox{\boldmath
$\hat{\eta}_{\overline{\nu}}$}\cdot{\bf \hat{q}}) - \gamma_{5}
S^{\prime\perp}_{\overline{\nu}}\cdot \gamma \right]\bigg\},
\eeq
where $S^{\prime\perp}_{\overline{\nu}} =
\left(0, \mbox{\boldmath $\eta_{\overline{\nu}} ^{\perp}$} =  \mbox{\boldmath
$\hat{\eta}_{\overline{\nu}}$} - (\mbox{\boldmath
$\hat{\eta}_{\overline{\nu}}$}\cdot{\bf \hat{q}}){\bf \hat{q}}\right)$. We see
that in spite of the singularities $m_{\overline{\nu}}^{-1}$ in the polarization four-vector
$S^{\prime}_{\overline{\nu}} $, the density operator $\Lambda_{\overline{\nu}}^{(s)}$ remains finite
including the transverse component of the antineutrino spin polarization
\cite{Michel}.

The corresponding formula for the  spin polarization 4-vector of  massive neutrino
$S^{\prime}_{\nu}$
%in its rest frame and for the initial neutrino
moving  with the momentum ${\bf q}$ is as follows:
\beq
S^{\prime}_{\nu} & = & (S^{\prime 0}_{\nu}, {\bf S^{\prime}}_{\nu}),\\
S^{\prime 0}_{\nu} & = & \frac{{|\bf q|}}{m_{\nu}}(\mbox{\boldmath
$\hat{\eta}_{\nu}$}\cdot{\bf \hat{q}}),
\\
{\bf S^{\prime}}_{\nu} & = &
\frac{E_{\nu}}{m_{\nu}}(\mbox{\boldmath
$\hat{\eta}_{\nu}$}\cdot{\bf \hat{q}}){\bf \hat{q}} +
\mbox{\boldmath $\hat{\eta}_{\nu}$} - (\mbox{\boldmath
$\hat{\eta}_{\nu}$}\cdot{\bf \hat{q}}){\bf \hat{q}}.
\eeq
The formula for the density
operator of the polarized neutrino in the limit of vanishing  neutrino
mass $m_{\nu} $  is given by:
\beq
\lim_{m_{\nu}\rightarrow 0}\Lambda_{\nu}^{(s)} &=&
\lim_{m_{\nu}\rightarrow 0}
\frac{1}{2}\bigg\{\left[(q^{\mu}\gamma_{\mu}) + m_{\nu}\right]\left[1 +
\gamma_{5}(S^{\prime \mu}_{\nu}\gamma_{\mu})\right]\bigg\}  \\
%& = & \mbox{} \frac{1}{2}\bigg\{(q^{\mu}\gamma_{\mu})
%\left[1 - \gamma_{5}(\mbox{\boldmath
%$\hat{\eta}_{\overline{\nu}}$}\cdot{\bf \hat{q}}) + \gamma_{5} (\mbox{\boldmath
%$\hat{\eta}_{\overline{\nu}}$} - (\mbox{\boldmath $\hat{\eta}_{\overline{\nu}}$}\cdot{\bf
%\hat{q}}){\bf \hat{q}})\cdot
%\mbox{\boldmath $\gamma$}\right]\bigg\} \\
 & = & \mbox{} \frac{1}{2}\bigg\{(q^{\mu}\gamma_{\mu})
 \left[1 + \gamma_{5}(\mbox{\boldmath
$\hat{\eta}_{\nu}$}\cdot{\bf \hat{q}}) + \gamma_{5}
S^{\prime\perp}_{\nu}\cdot \gamma \right]\bigg\}.
\eeq
\section{Antineutrino spectral function}
The formula for the electron antineutrino spectral function in case of the polarized muon decay at rest, when  the exotic $g_{LR}^{S}$ coupling of the right-chirality antineutrinos in addition to the standard $g_{LL}^{V}$ coupling of the left-chirality anineutrinos is admitted, takes the form:
\beq
   \label{didera}
 \frac{d^2 \Gamma}{ dy d\Omega_\nu}
 & = & \left(\frac{d^2 \Gamma}{ dy
d\Omega_\nu}\right)_{(V)} + \left(\frac{d^2 \Gamma}{dy
d\Omega_\nu}\right)_{(S )} + \left(\frac{d^2 \Gamma}{dy
d\Omega_\nu}\right)_{(VS)},
\\
\left(\frac{d^2 \Gamma}{ dy d\Omega_\nu}\right)_{(V)} & = & \frac{
G_{F}^2 m_{\mu}^5 }{128\pi^4}\bigg\{|g_{LL}^{V}|^2 
(1+\mbox{\boldmath $\hat{\eta}_{\overline{\nu}}$}\cdot\hat{\bf
q})(1+\mbox{\boldmath $\hat{\eta}_{\mu}$}\cdot\hat{\bf
q})y^2 (1 - y)\bigg\}, \label{livetimeV}
\\
\left(\frac{d^2 \Gamma}{ dy d\Omega_\nu}\right)_{(S)} & = & \frac{G_{F}^2
m_{\mu}^5}{3072\pi^4}|g_{LR}^{S}|^2 (1-\mbox{\boldmath
$\hat{\eta}_{\overline{\nu}}$}\cdot\hat{\bf q})
\nonumber \\
&& \mbox{} \cdot y^2 
\bigg\{(3 - 2y) - (1 - 2y)\mbox{\boldmath
$\hat{\eta}_{\mu}$}\cdot\hat{\bf q} 
\bigg\},
\label{livetimeS}
\eeq \noindent
 \beq\label{DDR}
 \left(\frac{d^2 \Gamma}{dy d\Omega_\nu}\right)_{(VS)} & = &
\frac{G_{F}^2 m_{\mu}^5 }{256\pi^4} \bigg\{ |\mbox{\boldmath
$\eta_{\overline{\nu}}^{\perp}$}| |\mbox{\boldmath $\eta_{\mu} ^{\perp}$}|
|g_{LL}^V ||g_{LR}^{S}| cos(\phi - \alpha_{VS}) \nonumber\\ && \mbox{} 
\cdot y^2(1- y)\bigg\},  \eeq
where $\mbox{\boldmath $\hat{\eta}_{\overline{\nu}}$}$,
$(\mbox{\boldmath$\hat{\eta}_{\overline{\nu}}$}\cdot\hat{\bf q}){\bf\hat{q}}$,
and $\mbox{\boldmath $\hat{\eta}_{\overline{\nu}}^{\perp}$}$ denote the unit
polarization vector, its longitudinal component, and transverse component of the
outgoing $\overline{\nu}_e$ in its rest system, respectively.
 $\mbox{\boldmath $\hat{\eta}_{\mu}$}$ is the unit polarization vector of the initial muon in its rest frame. 
  $y=\frac{2E_\nu}{m_\mu}$ is the reduced antineutrino energy
for the muon mass $m_\mu$, it varies from $0 $ to $1$, and
$d\Omega_\nu$ is the solid angle differential for $\overline{\nu}_e$
momentum $\hat{\bf q}$.\\
The interference term is presented for the case when  $\mbox{\boldmath $\hat{\eta}_{\mu}$} \cdot {\bf \hat{ q}}= 0$. $\phi$ is the angle between the $\mbox{\boldmath
$\eta_{\overline{\nu}}^{\perp}$}$ and the  $\mbox{\boldmath
$\eta_{\mu} ^{\perp}$}$, see Fig. 1.  
$\alpha_{V S} \equiv \alpha_{V}^{LL} - \alpha_{S}^{LR}$
 is the relative phase between the $g^{V}_{LL}$ and $g^{S}_{LR}$.
\\
With the use of the current data \cite{Data},  the upper limit on the
magnitude of the transverse antineutrino polarization and lower bound on the
longitudinal antineutrino polarization have been calculated, see \cite{Fetscher}:
\beq \label{trlo}
 |\mbox{\boldmath $\eta_{\overline{\nu} }^{\perp}$}|
&=& 2\sqrt{Q_{L}^{\overline{\nu}}(1-Q_{L}^{\overline{\nu}})}  \leq 0.088, \;
\mbox{\boldmath $\hat{\eta}_{\overline{\nu}}$}\cdot\hat{\bf q} = 2
Q_{L}^{\overline{\nu}} -1 \geq 0.996,
\\
Q_{L}^{\overline{\nu}} &=& 1 - \frac{1}{4}|g_{LR}^S|^{2} \geq
0.998,
\eeq
where $Q_{L}^{\overline{\nu}}$ is the probability of  the $\overline{\nu}_{e}$ to be 
left-chirality.

\end{document}